# Accelerated Assistant to Sub-Optimum Receiver for Multi Carrier Code Division Multiple Access System

Muhammad Adnan Khan[1]
Muhammad Umair[2]
Muhammad Aamer Saleem Choudry[3]
School of Engineering and Applied Sciences (SEAS)
ISRA University, Islamabad Campus
Islamabad, Pakistan
[1]adnan_600@yahoo.com, [2]umairbwp@gmail.com, [3]aamer.dr@gmail.com

*Abstract*— **The Multiple Input Multiple output (MIMO) system are considered to be the strongest candidate for the maximum utilization of available bandwidth. In this paper, the MIMO system with the combination of Multi-Carrier Code Division Multiple Access (MC-CDMA) and Space Time Coding (STC) using the Alamouti's scheme is considered. A Genetic Algorithm (GA) based receiver with an exceptional relationship between filter weights while detecting symbols is proposed. This scheme has better Convergence Rate and Bit Error Rate (BER) than the Fast- LMS Adaptive receiver.**

*Keywords—MIMO, MC-CDMA, STBC, GA, BER, Convergence Rate*

## I. INTRODUCTION

Multiple Input and Multiple Output (MIMO) system utilizes more than one antenna on both transmitters and receivers. This technology got good attention in 4th Generation (4G) wireless networks due to fact of maximum utilization of limited bandwidth [1].

The Alamouti's [2] Space Time Block Coding (STBC) is suggested for different flavours of Code Division Multiple Access (CDMA) based systems in [3], [4] and [5]. The Multi-Carrier Code Division Multiple Access (MC-CDMA) system is one of them. The MC-CDMA scheme is recommended for MIMO technology [5].

The Space Time Block Coding (STBC) is preferred for the Batch processing systems in [6], [7] and [8]. The batch processing is one of the affective methods in order to reduce the computation complexity of receiver. In batch processing system, the data is assembled in batch before computation begins. One of the key requirements of batch processing system is the calculation of inverse auto-correlation matrix of the received signal. The user configuration and channel co-efficient are not static. This increases the computational complexity. There are chances that filter vector length is large which burdens the receiver as it results in too many computations.

The adaptive implementation of such receiver is better approach in order to overcome the slow convergence rate and high Bit Error Rate (BER) problem. The Least Mean Square (LMS) and Fast Least Mean Square (Fast-LMS) adaptive receivers are proposed for MC-CDMA systems in [7] and [8]. The Fast-LMS is the improved version of [7]. In [8], a relationship based LMS adaptive receiver is proposed. This receiver has two times faster convergence rate than the proposed in [7].

In this paper, Genetic Algorithm (GA) based receiver using relationship in weights [8], [9] is proposed. Our proposed scheme has better convergence rate and Bit Error Rate (BER) than the schemes proposed in [8]. In GA, the breeding and mutation mechanisms are adopted in order to invent a population of individuals which is used to estimate the solution [10].

## II. MMSE BASED SYSTEM MODEL

We assumed Alamouti's STBC scheme based MC-CDMA system with two transmits antennas and one receive antenna. The number of transmits antenna can be increased as per future requirement.

It is needed to sent two successive symbols at each symbol interval due to Alamouti's scheme. The two successive symbols $b_u(2k-1)$ and $b_u(2k)$ are send from transmit antenna A and B at first symbol interval. The symbols $-b_u^*(2k)$ and $b_u^*(2k-1)$ are send from the transmit antenna A and B in the next symbol interval.

In CDMA based systems, the spreading codes are employed for frequency domain spreading. The couple $(c_{u,1}, c_{u,2})$ with dimensions $M \times 1$ is used for frequency domain spreading from transmit antenna A and transmit antenna B.

The multipath reflection from transmit antenna A and B towards single receive antenna C results in the delay spread. The ultimate delay spread is supposed to be inferior than the cyclic prefix length of all subcarriers. The frequency domain received signal due two successive symbols is $\mathbf{r}(k) = [r_1(k), ...., r_{2M}(k)]^T$. The FFT will reform it like [7]

$$\mathbf{r}(k) = \sum_{u=1}^{U} \{\mathbf{f}_{u,1} b_u(2k-1) + \mathbf{f}_{u,2} b_u(2k)\} + \mathbf{n}(k) \quad (1)$$

where

$U$ is Number of subscribers, $\mathbf{f}_{u,1} = [a_{u,1}^T \quad a_{u,2}^H]^T$ and $\mathbf{f}_{u,2} = [a_{u,2}^T \quad -a_{u,1}^H]^T$

where

$$\mathbf{a}_{u,p} = \mathbf{H}_{u,p} \mathbf{c}_{u,p}$$

where further $\mathbf{H}_{u,p} = diag(H_{u,p,0}, ...., H_{u,p,M-1})$ is the *uth* user channel response at *pth* transmit antenna. The $\mathbf{n}(k)$ is complex additive white Gaussian noise (AWGN) with zero mean and its covariance matrix is $\sigma_v^2 \mathbf{I}_{2M}$. The $\mathbf{I}_{2M}$ is the Identity matrix of size $2M \times 2M$.

The weight vectors $\mathbf{w}_1$ and $\mathbf{w}_2$ of filter with dimension $2M \times 1$ are needed in order to catch the $b_1(2k-1)$ and $b_1(2k)$. The filter output Mean Square Error (MSE) is defined as

$$C(\mathbf{w}_1, \mathbf{w}_2) = E[|\mathbf{w}_1^H r(k) - b_1(2k-1)|^2] +$$
$$E[|\mathbf{w}_2^H r(k) - b_1(2k)|^2] \quad (2)$$
$$= C(\mathbf{w}_1) + C(\mathbf{w}_2)$$

The minimization problem in [4] is required in order to attain the Minimum Mean Square Error (MMSE) receiver for the above mentioned STBC based MC-CDMA system:

$$[\mathbf{w}_{o,1}, \mathbf{w}_{o,2}] = \arg\min_{\mathbf{w}_1, \mathbf{w}_2} C(\mathbf{w}_1, \mathbf{w}_2)$$
$$= \{\min_{\mathbf{w}_1} C(\mathbf{w}_1) + \min_{\mathbf{w}_2} C(\mathbf{w}_2)\} \quad (3)$$

The MMSE receiver works by setting the derivatives of these filter weight $\mathbf{w}_1$ and $\mathbf{w}_2$ to zero which is

$$\mathbf{w}_{o,1} = \mathbf{R}_y^{-1} \mathbf{f}_{1,1}, \quad \mathbf{w}_{o,2} = \mathbf{R}_y^{-1} \mathbf{f}_{1,2} \quad (4)$$

Where $\mathbf{R}_y = E[\mathbf{r}(k)\mathbf{r}^H(k)]$ which is the autocorrelation matrix of $\mathbf{r}(k)$. The MMSE with respect to (4) is

$$C_{\min} = C(\mathbf{w}_{o,1}, \mathbf{w}_{o,2})$$
$$= (1 - \mathbf{f}_{1,1}^H \mathbf{R}_y^{-1} \mathbf{f}_{1,1}) + (1 - \mathbf{f}_{1,2}^H \mathbf{R}_y^{-1} \mathbf{f}_{1,2}) \quad (5)$$

## III. ACCELERATED RECEIVER PROPERTIES

### A. Relationship between filter weight vectors of MMSE

The unique association between the optimum weights $\mathbf{w}_{o,1}$ and $\mathbf{w}_{o,2}$ is described in this part as in [8]. Suppose $\mathbf{R}_y$ by its sub matrices of dimensions $M \times M$:

$$\mathbf{R}_y = \begin{bmatrix} \mathbf{R}_a & \mathbf{R}_b \\ \mathbf{R}_c & \mathbf{R}_d \end{bmatrix} \quad (6)$$

It was noted in that $\mathbf{R}_y$ has a certain relationship in its diagonals. The relationship is

$$\mathbf{R}_d = \mathbf{R}_a^* \text{ and } \mathbf{R}_c = -\mathbf{R}_b^* \quad (7)$$

The association found in (7) between MMSE filter weight vectors is used in (4) to derive a special relationship. The optimal weight vectors $\mathbf{w}_{o,1}$ and $\mathbf{w}_{o,2}$ of dimension $M \times 1$ are presented as

$$\mathbf{w}_{o,1} = \begin{bmatrix} \mathbf{w}_{o,a} \\ \mathbf{w}_{o,b} \end{bmatrix}, \quad \mathbf{w}_{o,2} = \begin{bmatrix} \mathbf{w}_{o,c} \\ \mathbf{w}_{o,d} \end{bmatrix} \quad (8)$$

and the following relation is satisfied by these vectors:

$$\mathbf{w}_{o,b} = \mathbf{w}_{o,c}^*, \quad \mathbf{w}_{o,d} = -\mathbf{w}_{o,a}^* \quad (9)$$

### B. Cost function

In this part, the MMSE cost function is presented. The MMSE filter weight vectors are given in (8) satisfies the relation given in (9). So it's a very good sign in order to increase the convergence rate by only updating the weight vectors satisfying the relationship in (9) as in [8]. The cost function of MSE in (2) can be reformed as only function of $\mathbf{w}_a$ and $\mathbf{w}_c$:

$$C_N(\mathbf{w}_a, \mathbf{w}_c) = C_{N1}(\mathbf{w}_a, \mathbf{w}_c) + C_{N2}(\mathbf{w}_a, \mathbf{w}_c) \quad (10)$$

Where

$$C_{N1}(\mathbf{w}_a, \mathbf{w}_c) =$$
$$E[|\mathbf{w}_a^H r(2k-1) + \mathbf{w}_c^T r^*(2k) - b_1(2k-1)|^2]$$

and

$$C_{N2}(\mathbf{w}_a, \mathbf{w}_c) =$$
$$E[|\mathbf{w}_c^H r(2k-1) + \mathbf{w}_a^T r^*(2k) - b_1(2k)|^2]$$

## IV. GA EXPLANATION

There are many algorithms that are being used for optimization problems. The Genetic Algorithm (GA) is one of the very frequently used optimization Algorithm. The GA works effectively on the nonlinear problems.

In GA, the solution is represented as chromosome. The new population is always generated by breeding and mutation mechanisms [9]. The number of weights required for solution is $M$ instead of $2M$. The other $M$ is calculated for the relationship given in (9). Each column of weight matrix is considered to be a gene. The total number of columns of this matrix is $M_T$. The $q$th element weight vector is given by:

$$\mathbf{w}_q = [w_{q,1}, w_{q,2}, ..., w_{q,M_G}]$$

In our proposed receiver, we have applied the Genetic Algorithm (GA) in order to minimum of cost function given in (10). The GA description is given below in Table 1.

The Table 1 is the algorithm explaining GA. Initially, a population of $q$ individuals $\mathbf{W} = [\overline{w}_1, \overline{w}_2]$ is created randomly. The recursive execution process mechanism is used in GA after initialization of population. One generation is produced on each recursive cycle. The two individuals are picked at random which are to be the parents of next generation. These parents give birth to two children. The fitness function is given in (10) for calculation of fitness of children. Further a new population is produced with the genes giving the best answers with fitness function given in (10). We continue the process till a threshold value is achieved or certain numbers of cycles are completed.

Let the two parents are $[p_1, p_2]$. We analyzed four different approaches for parent variety. These are: Eugenic selection, Alpha-male selection, preferred selection and Random selection. In Eugenic selection, the two best parents $p_1 = w_1$ and $p_2 = w_2$ are selected on each cycle. In next approach of Alpha-male selection, the first is the best individual $p_1 = w_1$ and second one is selected randomly from the remaining population $p_2 \in \{w_k, 2 < k \leq K\}$. In preferred selection, $p_1$ is selected from $p_1 \in \{w_j, 1 < j < k\}$ excluding finest ones and $p_2$ is selected randomly from the outstanding population $p_2 \in \{w_k, 2 < k < K\}$ superior than $p_1$. In last strategy of random selection, $p_1$ and $p_2$ are selected randomly at each cycle. We find out that the preferred selection gives the superior result in our analysis. The two selected parents produce two children $c_1$ and $c_2$. The $c_1$ and $c_2$ have same genes as parents altogether. Any of the child $c_1$ or $c_2$ enclose number of genes from parent $p_1$ and outstanding genes from $p_2$.

We observed two crossover ratios: ½ and ½ and ¾ and ¼. The crossover of parents with the ratios of ¾ and ¼ gives good result in our case. We choose $N$ individuals that give us the minimized cost function. We choose further one random index of both column and row. We changed the sign of bit placed in that location. This new generation is replaced by

$$w_1 = \arg\min_{x \in \chi} C_{N1}(x)$$
$$w_2 = \arg\min_{x \in \chi} C_{N2}(x) \quad (11)$$

This new generation replaces the old generation of W. The population is rearranged in mounting order of cost after the population has been reorganized. The algorithm moves on next generation or ends if stopping condition is achieved or number of cycles are ended.

TABLE I. GENETIC ALGORITHM (GA) DESCRIPTION

| GA Algorithm | |
|---|---|
| **S. No** | **Steps** |
| 1. | Start |
| 2. | Initialization of weights $\mathbf{w}_q$ |
| 3. | Calculate the fitness function using the cost function given in (10) |
| 4. | Sort the weights in ascending order as per fitness values. |
| 5. | Select the best parents |
| 6. | Generate the children using the crossover. (The crossover ration is ¾ and ¼ of Parent.No.1 and Parent.No.2) |
| 7. | Mutation process is applied |
| 8. | Calculate the fitness |
| 9. | If (Number of cycles) go to step 10 Else go to step 3 |
| 10. | Stop |

V. RESULTS

All results are simulation based using the MATLAB R2009a. The uplink MC-CDMA system is implemented with $M = 32$ subcarriers and number of users is $K = 20$. The subcarriers are equal to the dimension of the spreading code. The real and imaginary parts of the spreading code are independently chosen from $1/\sqrt{2}$ and $-1/\sqrt{2}$ at random. The weights to be optimized are complex. The Rayleigh fading channel is implemented using three paths. In all cycles, the channel coefficients as well as spreading code sequence is constant.

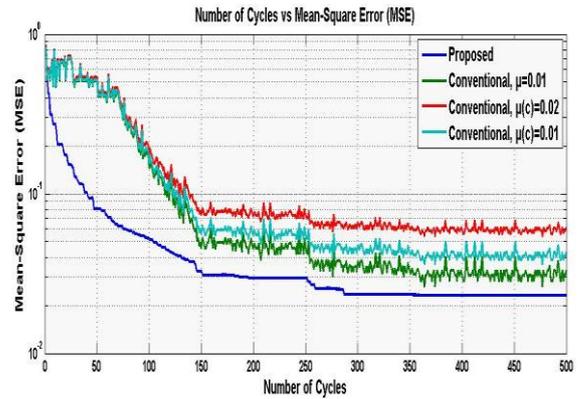

Figure 1. Number of Cycles Vs Mean Square Error (**MSE**)

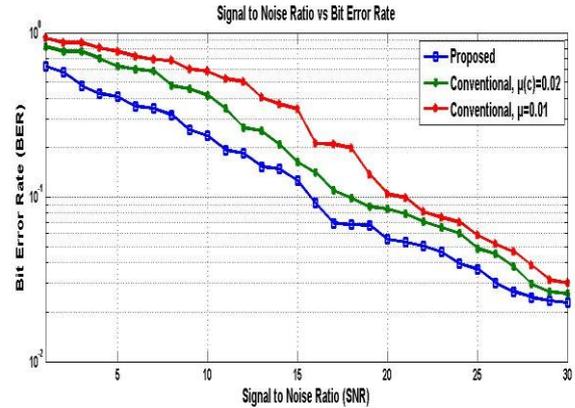

Figure 2. Signal to Noise Ratio (**SNR**) Vs Bit Error Rate (**BER**)

The Figure 1 shows the Mean Square Error (MSE) with respect to Number of Cycles (NOC). The low most curve represent the proposed GA based Sub-Optimum accelerated receiver. The remaining three curves are of conventional schemes in [7] and [8] with $\mu = 0.01, \mu_c = 0.02$ and $\mu_c = 0.01$ from top to down. The MSE of proposed Accelerated receiver is low as compare to conventional schemes at initial cycles. It is seem that the MSE of proposed scheme is better over all simulation cycles. It is also observed that the convergence rate is also faster with respect to the other conventional schemes.

The Figure 2 shows the filter output Bit Error Rate (BER) averaged over every two consecutive symbols while varying the SNR. The proposed sub optimum receiver BER is represented by the bottom most line in the figure 2. The second line from the bottom represents the conventional receiver with $\mu(c) = 0.01$. The topmost line represents the conventional receiver $\mu(c) = 0.02$. The proposed scheme is better at low and high SNR as its BER is low than conventional schemes.

## VI. CONCLUSION

We adopted GA with exceptional relationship between weights is used to optimize the uplink MC-CDMA receiver with Alamouti's STBC. The GA is applied in order to optimize the weights of MMSE receiver. The convergence rate and BER is studied for the proposed scheme. The convergence rate is faster and BER is also less than the schemes proposed in [7] and [8].